\documentstyle[aps,preprint,epsf]{revtex}
\tighten
\begin{document}
\draft
\title{Elastic scattering and breakup of $^{17}$F at 10 MeV/nucleon}
\author{J.~F.~Liang$^{1}$, J.~R.~Beene$^{1}$, 
H.~Esbensen$^{2}$, A.~Galindo-Uribarri$^{1}$, 
J.~Gomez~del~Campo$^{1}$, C.~J.~Gross$^{1,3}$, M.~L.~Halbert$^{1}$,
P.~E.~Mueller$^{1}$, D.~Shapira$^{1}$, D.~W.~Stracener$^{1}$, 
I.~J.~Thompson$^{4}$ and R.~L.~Varner$^{1}$}
\address{$^{1}$Physics Division, Oak Ridge National Laboratory, Oak Ridge, TN 
37831, USA}
\address{$^{2}$Physics Division, Argonne National Laboratory, Argonne, IL
60439, USA}
\address{$^{3}$Oak Ridge Institute for Science and Education, Oak Ridge, TN 
37831, USA}
\address{$^{4}$Department of Physics, University of Surrey, Guildford GU2 5XH,
United Kingdom}
\date{\today}
\maketitle
\begin{abstract}
Angular distributions of fluorine and oxygen produced from
170 MeV $^{17}$F incident on $^{208}$Pb were measured. The elastic scattering 
data are in good agreement with optical model calculations using a
double-folding potential and parameters similar to those obtained from 
$^{16}$O+$^{208}$Pb. A large yield of oxygen was observed near $\theta_{\rm
lab}=36^{\circ}$.
It is reproduced fairly well by a calculation of the ($^{17}$F,$^{16}$O) 
breakup, which is dominated by one-proton stripping reactions. The discrepancy
between our previous coincidence measurement and theoretical predictions
was resolved by including core absorption in the present calculation.
\end{abstract}

Our knowledge of atomic nuclei comes from experiments with nuclei in the 
valley of stability. The advent of radioactive beams enables us to
explore nuclei near and beyond the drip lines and provides tests of 
current nuclear structure models. Nuclear
reactions are often used as tools for studying nuclear
structure. Breakup is an important reaction channel in the scattering of 
weakly bound nuclei and can be a rich source of
information on reaction mechanisms and the
structure of such nuclei\cite{an93,he96,ba96,na98}.

Fluorine-17 is a proton drip-line nucleus with its valence proton bound by
0.6 MeV. Because of this loose binding, the r.m.s. radius, 
3.7 fm\cite{mo97}, is significantly larger than that of the $^{16}$O core. 
Furthermore, 
the first excited state of $^{17}$F, E$_{\rm x}$ = 0.495 MeV and J$^{\pi}$ = 
$\frac{1}{2}^{+}$, is reported to have an extended 
r.m.s. radius, 5.3 fm\cite{mo97}, and is considered to be a nuclear halo state. 
Studies of the influence on fusion of breakup of weakly bound nuclei has
generated considerable interest in recent years\cite{si97}.
The measurement of fusion of $^{17}$F on $^{208}$Pb at energies near the
Coulomb barrier found no fusion enhancement\cite{re98}. In our previous
measurement of the 
breakup of $^{17}$F by detecting the breakup products, proton and $^{16}$O,
simultaneously, a very
small cross section was observed and disagreed with theoretical 
predictions\cite{li00}. That experiment used a large area 
detector to optimize the collection efficiency;
the measured cross sections were averaged over the angles spanned by the
detector.

A recent study of the
breakup reaction $^{8}$B $\rightarrow$ $^{7}$Be+p on a $^{58}$Ni target 
at sub-Coulomb energies~\cite{gu00} demonstrated that the resulting $^{7}$Be
angular distribution can be accounted for by two different theoretical 
approaches, coupled discretized-continuum channels (CDCC)\cite{nu99} and
dynamical calculation\cite{es99}. With this in mind, we have performed a new 
experiment on the breakup of $^{17}$F by 
measuring the angular distribution of reaction products in singles. The data 
provide some details of the cross section over a range of angles and no Monte
Carlo simulation is required for obtaining the efficiency of the detectors,
unlike the coincidence measurement.
However, the identification of reaction products with the breakup channel 
is not as straightforward as in the coincidence measurement, since
similar products can be produced in several reaction channels. 
With the aid of theoretical calculations, the
dominant reaction products were shown to originate from breakup.

The experiment was carried out at the Holifield Radioactive Ion Beam Facility
(HRIBF) using a 170 MeV $^{17}$F beam incident on a 2 mg/cm$^{2}$ $^{208}$Pb 
target. A 44 MeV deuteron beam provided by the Oak Ridge Isochronous Cyclotron
(ORIC) was used to bombard a fibrous hafnium oxide target to produce 
radioactive $^{17}$F by the $^{16}$O(d,n) reaction\cite{we99}. The A=17 ions
were extracted from the target-ion source, mass analyzed and
subsequently accelerated by the 25 MV tandem postaccelerator. The $^{17}$O 
isobar contaminant was removed by inserting a carbon stripper
foil at the exit of the tandem accelerator, before the 90$^{\circ}$
analyzing magnet, and selecting the fully stripped $^{17}$F$^{9+}$ ion beam. 
The beam was monitored by a Si surface barrier detector positioned at 
10$^{\circ}$ to the beam direction in the target chamber as well as by 
the focal plane detector in the Enge spectrograph at 3$^{\circ}$ on the other 
side of the beam. The average beam intensity was $5 \times 10^{5}$ $^{17}$F/s. 
The reaction 
products were detected at $\theta_{\rm lab}=45^{\circ}$ by a 
$\Delta$E-E telescope consisting of a large area (900~mm$^{2}$) 
100~$\mu$m Si surface barrier 
detector (SBD) and a 5~cm$\times$5~cm, 300~$\mu$m
double-sided Si strip detector (DSSD) behind the SBD.
The DSSD, which has 16 vertical
and 16 horizontal strips, was placed 8.3~cm from the target, resulting in an 
angular resolution of $\sim~2^{\circ}$. At forward 
angles, the elastically scattered $^{17}$F was measured by the DSSD only. 

Figure~\ref{fg:ede} shows the E vs. $\Delta$E histogram  of the SBD-DSSD
telescope for events in all the strips.
Reaction products of Z=9(F), 8(O), 7(N), 6(C), 5(B) and 1(H) were observed.
The band of constant energy loss ($\Delta$E) at channel number 
$\sim$~135, corresponding to the energy loss of the elastically scattered
$^{17}$F, results from the positron decay of $^{17}$F stopped in the DSSD. 
These events were recorded when a positron
emission took place at the same time as an elastic $^{17}$F
struck another strip in the DSSD.

The elastic scattering data were extracted from the Z=9 products identified in 
this E vs. $\Delta$E histogram.
The energy resolution of the detectors was not good enough to allow a 
clear separation
between the elastic and inelastic scattering events. Moreover,
the mass of the reaction products cannot be identified in this experiment.
Since there are some neutron pickup channels with positive Q values,
the elastic scattering data may include contributions from inelastic
scattering and neutron transfer reactions.
The absolute cross section was obtained by normalizing the yields 
to the Si detector at 10$^{\circ}$ where the elastic scattering 
was taken as Rutherford scattering.
The elastic scattering angular distribution is shown in Fig.~\ref{fg:elas1}.
As can be seen, the assumption of pure Rutherford scattering at 10$^{\circ}$ 
is valid.

Optical model fits to the elastic scattering data were performed using the 
code {\tt Ptolemy}\cite{ma78}. The Woods-Saxon potential parameters were taken 
from the 192 MeV $^{16}$O + 
$^{208}$Pb elastic scattering\cite{ba75} and only the depth of the imaginary 
potential was allowed to vary. The best-fit result is shown by the dotted
curve in Fig.~\ref{fg:elas1} and the parameters are listed
in Table~\ref{tb:opm} as SET~I. Parameters in SET~II of 
Table~\ref{tb:opm} were obtained using the procedures described in 
Ref.~\cite{ba75} by fitting the data with the depth of the real
potential, V, fixed at 40 MeV and varying the other parameters. Throughout the 
fitting processes, the shape of the 
real and imaginary potential was assumed to be the same. 
The imaginary potential depth, W, was changed from 35 to 95~MeV in 
5~MeV steps and the radius and diffuseness parameters, r and a, were varied 
to find the minimum $\chi^{2}$. The best fit is shown by the dash-dotted curve. 
The radius and diffuseness parameters were 
very similar to those obtained in the 140 MeV $^{16}$O + $^{208}$Pb elastic
scattering\cite{ko73}.
The dashed curve, corresponding to parameter SET~III, 
is the result of allowing all the parameters, V, W, r, and a, to vary subject 
to the constraint that r and a were identical for the real and imaginary
potentials. The dashed curve and dash-dotted curve are almost 
indistinguishable. All three sets of parameters describe the data very 
well at 
angles smaller than 41$^{\circ}$. However, differences between the dotted 
curve and dashed curve can be seen at larger angles. The inclusion of 
inelastic scattering and transfer in the data may account for the fact that
the experimental points are above the dotted prediction.

The elastic scattering data were also compared to calculations using 
potentials obtained from a double-folding model. The 
effective nucleon-nucleon interaction was taken from Satchler's 
systematics\cite{sa94} which uses the three-parameter Yukawa form 
\[ {\rm U_{NN}=-(V+iW)\frac{e^{-s/t}}{s/t} } \]
where s is the distance between the two nucleons and t = 0.7 fm. The 
depth of the potentials was ${\rm V=W=60-0.3E/A=57}$ MeV where E/A 
is the energy per nucleon of the projectile. The nuclear density 
distribution of $^{208}$Pb was constructed using the two-parameter Fermi form 
with parameters given in Table~2 of Ref.~\cite{sa94}. 
The ground state density distribution of $^{17}$F was constructed from shell 
model single particle wave functions as described by Satchler\cite{sa79}
(method B). 
Calculations using this double-folding potential with no adjustable parameters
are in very good agreement with the data, as shown by 
the solid curve in Fig.~\ref{fg:elas2}. Beyond about 45$^{\circ}$, the 
calculation underpredicts the data somewhat, which may 
be due to the inclusion of inelastic
scattering and neutron transfer reactions in the data. The inelastic 
excitation of $^{17}$F to its first excited state, E$_{\rm x}$ = 0.495 MeV and 
J$^{\pi}=\frac{1}{2}^{+}$, and $^{208}$Pb to its first 2$^{+}$,
E$_{\rm x}$ = 4.09 MeV, and 3$^{-}$, E$_{\em x}$ = 2.61 MeV, 
states were calculated in the
DWBA by the code {\tt Ptolemy}. The sum of the inelastic scattering
predicted by DWBA and the elastic scattering predicted by the folding 
potential is shown by the dotted curve in Fig.~\ref{fg:elas2} which
is in excellent agreement with the data.
Among the three inelastic
channels, the excitation cross section for populating the first excited state 
of $^{17}$F is calculated to be the largest, followed by exciting $^{208}$Pb 
to its first 3$^{-}$ state. Since the single particle wave function used seems
to give a good account of the ground state density distribution of $^{17}$F,
the halo property of the first excited state in $^{17}$F could be studied if
the inelastic scattering could be resolved in the data.
Coupled-channels calculations were performed with these three
inelastic channels using {\tt Ptolemy}. As can be seen by the dashed curve
in Fig.~\ref{fg:elas2}, the elastic scattering is influenced 
only slightly by the channel couplings. The one-neutron
transfer reaction was calculated by DWBA using {\tt Ptolemy}. The transfer
yield was predicted to be at least two orders of magnitude less than the 
elastic scattering and can be ignored, consistent
with the agreement between the data and the dotted curve in 
Fig.~\ref{fg:elas2}.

The angular distribution of Z=8 reaction products used the same normalization
as the elastic scattering.
Since this experiment provided no mass identification of the reaction products,
the observed oxygen
can originate from nucleon transfer, charge exchange, and breakup.
The one-proton transfer has the largest Q value of all transfer channels
leading to an oxygen isotope, $^{16}$O, as the final product.
Finite range DWBA calculations 
using the code {\tt Ptolemy} were performed to predict the cross section
angular distribution of
$^{208}$Pb($^{17}$F,$^{16}$O)$^{209}$Bi. Transfer to the lowest six single
particle states in $^{209}$Bi was calculated. The spectroscopic factors
were assumed to be 1 which will give upper limits for the cross sections.
The sum of the six cross sections is shown by the dotted curve in
Fig.~\ref{fg:strp}. It can be seen that the one-proton transfer cannot account 
for the yields of the oxygen angular distribution. Similar results were 
obtained for calculations using the code {\tt FRESCO}\cite{th88}. 
Although reactions leading to $^{18}$O and $^{207}$Bi in the exit channel have 
positive Q values, they cannot occur by simple
single-step transfer processes. Therefore, the cross sections are
expected to be smaller than that of one-proton transfer\cite{ko74}.

Oxygen-17 can be produced by the charge exchange reaction, 
$^{208}$Pb($^{17}$F,$^{17}$O)$^{208}$Bi, with a Q value of --0.11 MeV. 
Calculations were performed to estimate the cross section for this
reaction channel. Charge exchange reactions can take place by direct charge
exchange, a combination of inelastic excitation and direct charge exchange,
or by successive one-nucleon exchange. Since the direct charge exchange is 
orders of magnitude smaller than one-nucleon transfer\cite{du74}, calculations
were carried out using {\tt FRESCO} to model 
the successive one-nucleon exchange 
processes\cite{ke01}. Two possible modes, proton stripping 
followed by neutron pickup and neutron pickup followed by proton stripping,
were considered. Only transfer between ground states was calculated. The
charge exchange cross section is calculated to be 2 orders of magnitude
smaller than that of one-proton transfer calculated by DWBA. Measurements of
$^{28}$Si($^{18}$O,$^{18}$F)$^{28}$Al show that the charge exchange 
cross section is at least an order of magnitude smaller
than that of one-nucleon transfer\cite{ki79}, consistent with
the results of our calculations. Since DWBA calculations predict 
that the one-proton transfer cross sections are much smaller than the 
measured cross sections, it is safe to assume that charge exchange can be 
ignored. 

The measurement is inclusive, so products from
two breakup processes, diffraction and stripping, can contribute to the
data\cite{he96}. In diffraction dissociation, 
the projectile breaks up, leaving the valence
nucleon in the continuum and the core intact. The final
state of the reaction consists of the valence nucleon, the core of the 
projectile and the target nucleus in its ground state. This is the process 
that our previous 
coincidence experiment measured\cite{li00}. The stripping
breakup depends on the core-target and nucleon-target absorption potentials.
In a loosely bound nuclear system, if the separation energy (S$_{\rm N}$) 
is much less than the kinetic energy per nucleon, S$_{\rm N} <<$ E/A, the
valence nucleon and the core can be treated approximately as independent 
particles\cite{ya92}. The effects of nucleon-target 
and core-target interactions become noticeable if the interactions are strong.

Semiclassical calculations of breakup similar to those in 
Ref.~\cite{he96} were performed to investigate these processes. The wave 
function of 
the projectile after interacting with the target is described by the ground
state wave function of the projectile and the profile functions of the proton
and $^{16}$O core. The interaction between the target nucleus and the 
constituents of the projectile was obtained from Ref.~\cite{va91}. 
The $^{16}$O-target optical potential was calculated in a single-folding model
based on the CH89 optical model potential\cite{va91}. The breakup probability 
as a function
of the impact parameter was calculated and converted to an angular distribution
assuming pure Coulomb scattering. The calculated angular 
distribution was converted from the center of mass of the reaction to the
laboratory frame using the Jacobian for elastic scattering. Because 
the reaction energy is high, the uncertainty introduced by
this approximation is smaller than the angular resolution of the DSSD.

Figure~\ref{fg:strp} presents the comparison of the experimental oxygen cross
section data with model calculations. The long-dashed curve is the
prediction for stripping breakup while the short-dashed curve is for
diffraction breakup. Apparently, stripping breakup is the
dominant reaction in the measured angular range. The solid curve shows the 
sum of diffraction and stripping.
It overpredicts the measured cross sections slightly; however,
the data do not extend far enough forward for a good comparison.
It should be pointed out that the calculations employed here did not consider
the recoil of the $^{16}$O. In addition, excitation of the $^{16}$O and
$^{208}$Pb in the final state was ignored. Measurements of the breakup
of $^{17}$C by $^{197}$Au show that core excitation is important for
nuclear breakup\cite{ma01}, but very small for Coulomb breakup\cite{sh01}. 
In our case, the 
measurement was performed near the grazing angle where the nuclear breakup 
should be significant. Therefore, the excitation of the core and target nuclei
should be considered. Furthermore, the stripping and
diffraction breakup were calculated separately. A more complete theoretical
treatment of this subject is underway and will be published 
elsewhere\cite{he01}.

Diffraction breakup of $^{17}$F was measured previously in our coincidence
experiment\cite{li00}. A first-order perturbation calculation overpredicted
the measured cross section by a factor of 4. The present work shows
the importance of stripping breakup resulting from the nucleon-target 
interaction. The discrepancy in the coincidence experiment can be
accounted for by including the core absorption by the target nucleus which 
was not
considered in the previous calculations. As shown in Fig.~\ref{fg:dfr},
the data agree with the calculations taking into account the core-target
interaction (short-dashed curve). The calculation is also consistent 
with predictions made by the CDCC calculation\cite{nu99} shown by the 
dash-dotted curve. The CDCC calculation takes into account
of higher order effects, continuum-continuum couplings, and full three-body
kinematics to produce results in the laboratory frame\cite{to01}.

It is shown in Ref.~\cite{he96} that the breakup of $^{11}$Be 
(S$_{\rm N} = 0.5$ MeV) by $^{208}$Pb
is dominated by stripping at 800 MeV/nucleon. At 40 MeV/nucleon, the stripping 
and diffraction breakup cross sections become similar~\cite{he96}. In the 
breakup of $^{8}$B at energies below the
Coulomb barrier, the stripping and diffraction breakup cross sections are also
similar\cite{gu00}. In this work, it is found that
the stripping breakup cross section of 10 MeV/nucleon $^{17}$F is almost 
twice as large as that of the diffraction breakup. 

In summary,
the elastic scattering and breakup of $^{17}$F was measured on a $^{208}$Pb
target. The
elastic scattering data were fitted with an optical model using a Woods-Saxon
potential. The potential parameters are very similar to those obtained
from $^{16}$O+$^{208}$Pb elastic scattering. The potential generated
from a double-folding model using Satchler's systematics also reproduced
the elastic scattering data. The dominant contribution to the measured angular 
distribution of oxygen nuclei in the exit channel was found to be the 
one-proton stripping reaction resulting in $^{16}$O. 
This demonstrates the importance of the proton-target
interaction in the breakup of $^{17}$F on $^{208}$Pb. Based 
on this finding, calculations of diffraction breakup considering core 
absorption were carried out. The new calculations agree with our previous
coincidence measurement.

We would like to thank J.~B.~Ball, D.~J.~Dean, J.~J.~Kolata, F.~M.~Nunes,
G.~R.~Satchler, and C.~Y.~Wong for informative discussions. We are specially
grateful to N. Keeley for the charge exchange calculations. 
The experimental measurements would not have been possible without the hard 
work of the HRIBF staff. Research at the Oak Ridge National Laboratory is 
supported by the U.S. Department of Energy under contract DE-AC05-00OR22725 
with UT-Battelle, LLC. The ORISE is supported by the U.S. Department
of Energy under contract number DE-AC05-00OR22750. One of us (H.E.) was
supported by the U.S. Department of Energy, Nuclear Physics Division, under
Contract No. W-31-109-ENG-38.

\begin{table}
\begin{tabular}{ccccc}
	& V 	& W     & r = r$_{\rm i}$    	& a = a$_{\rm i}$   \\ 
	& (MeV)	& (MeV)	& (fm)			& (fm)		\\ \hline
SET I	& 40  	& 49    & 1.23			& 0.63          \\ 
SET II	& 40 	& 75	& 1.28			& 0.50		\\
SET III	& 50 	& 93	& 1.27			& 0.51		\\
\end{tabular}
\caption{Optical model potential parameters obtained from fitting the 170 MeV
$^{17}$F+$^{208}$Pb elastic scattering data. The 
depth of the real and imaginary potentials is shown by V and W, respectively.
The real and imaginary radius parameters are shown by r and r$_{\rm i}$,
respectively. The real and imaginary diffuseness parameters are shown by
a and a$_{\rm i}$, respectively.
\label{tb:opm}}
\end{table}

\begin{figure}
\caption{Histogram of E vs. $\Delta$E obtained by the SBD-DSSD telescope.
\label{fg:ede}}
\end{figure}

\begin{figure}
\caption{Angular distribution of the ratio of measured elastic scattering 
to calculated Rutherford scattering.
The optical potential fit to the data is shown by 
the dotted curve for parameters SET~I, dash-dotted curve for SET~II, and
dashed curve for SET~III. The dash-dotted curve and dashed curve are almost
indistinguishable.
\label{fg:elas1}}
\end{figure}

\begin{figure}
\caption{Angular distribution of the ratio of measured elastic scattering 
to calculated Rutherford scattering.
The solid curve is the result of calculations using the optical 
potential obtained from a double-folding model. The sum of elastic and 
inelastic scattering calculations is shown by the dotted curve. The result of
coupled-channels calculations is shown by the dashed curve.
\label{fg:elas2}}
\end{figure}

\begin{figure}
\caption{Angular distribution of oxygen. The long-dashed and short-dashed
curves are for stripping and diffraction breakup, respectively.
The dotted curve is the result of a one-proton transfer DWBA calculation.
The solid curve represents the sum of stripping and diffraction breakup.} 
\label{fg:strp}
\end{figure}

\begin{figure}
\caption{Data and predictions of diffraction breakup. The long-dashed curve 
is for the calculation
without considering core absorption and the short-dashed curve includes
core absorption. The CDCC calculation is shown by the dash-dotted curve. The
open point is from our previous measurement\protect\cite{li00}.}
\label{fg:dfr}
\end{figure}

\newpage
\begin{figure}
\epsfxsize=6.0in
\epsffile{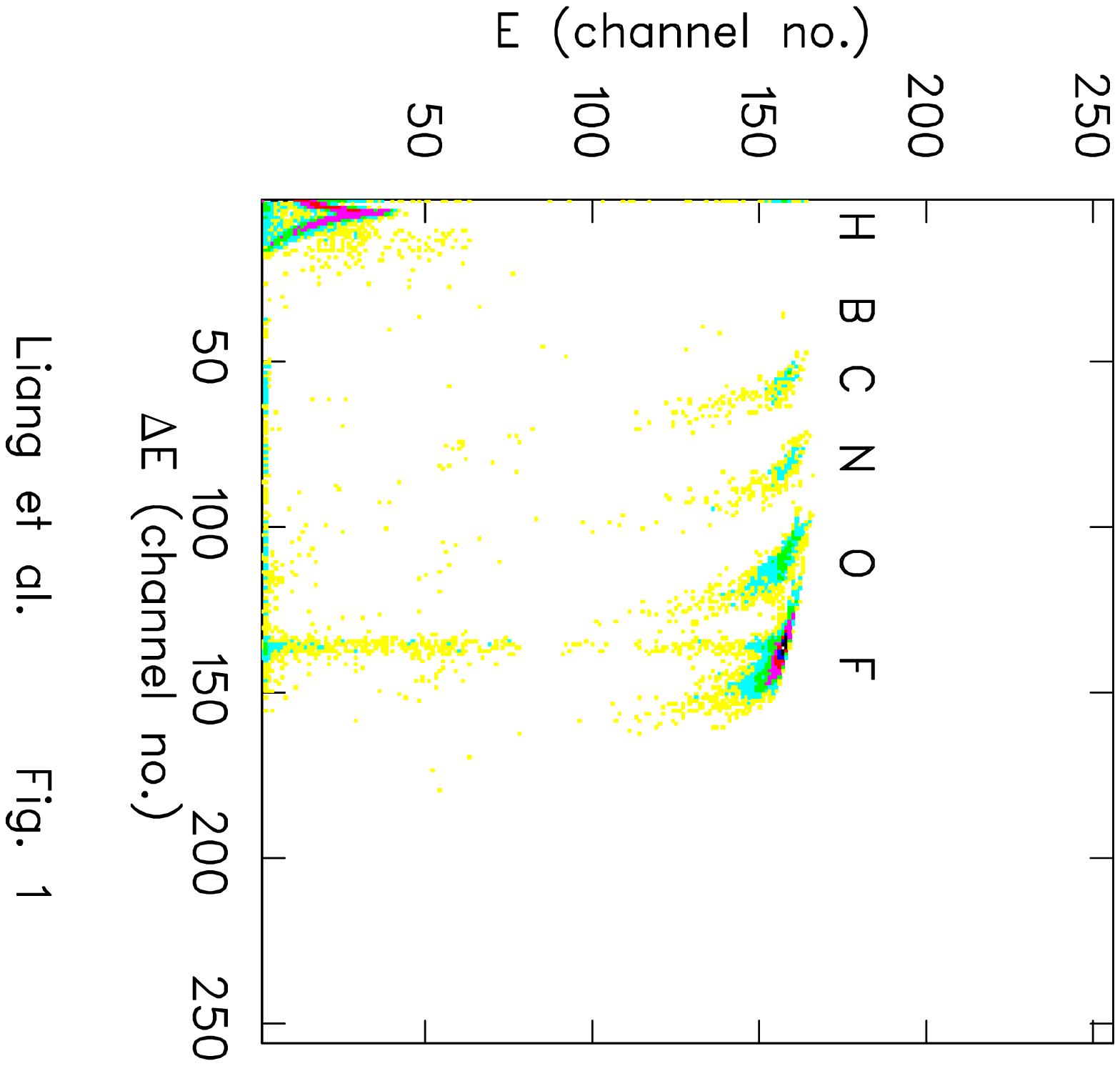}
\end{figure}
\begin{figure}
\epsfxsize=6.0in
\epsffile{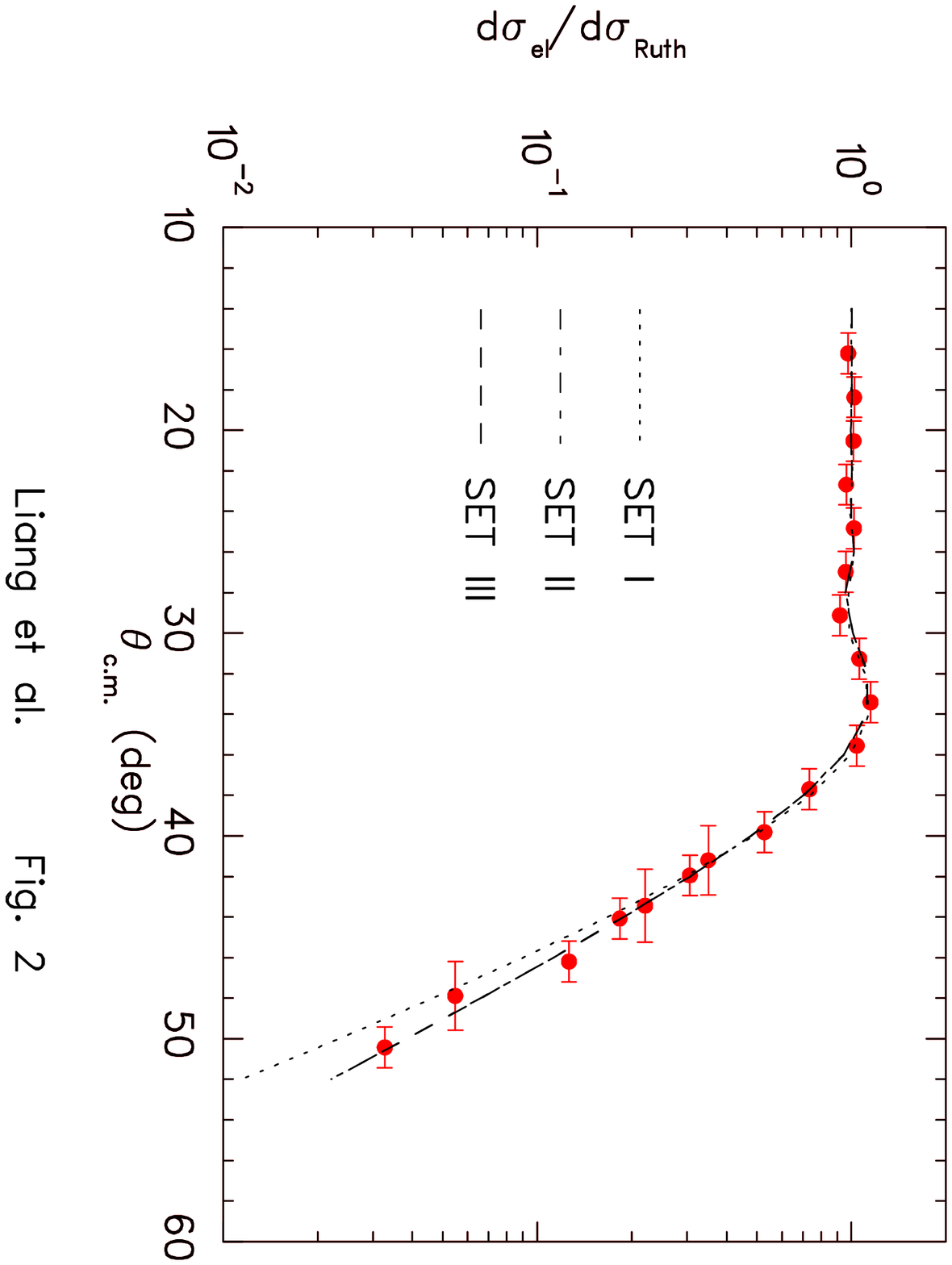}
\end{figure}
\begin{figure}
\epsfxsize=6.0in
\epsffile{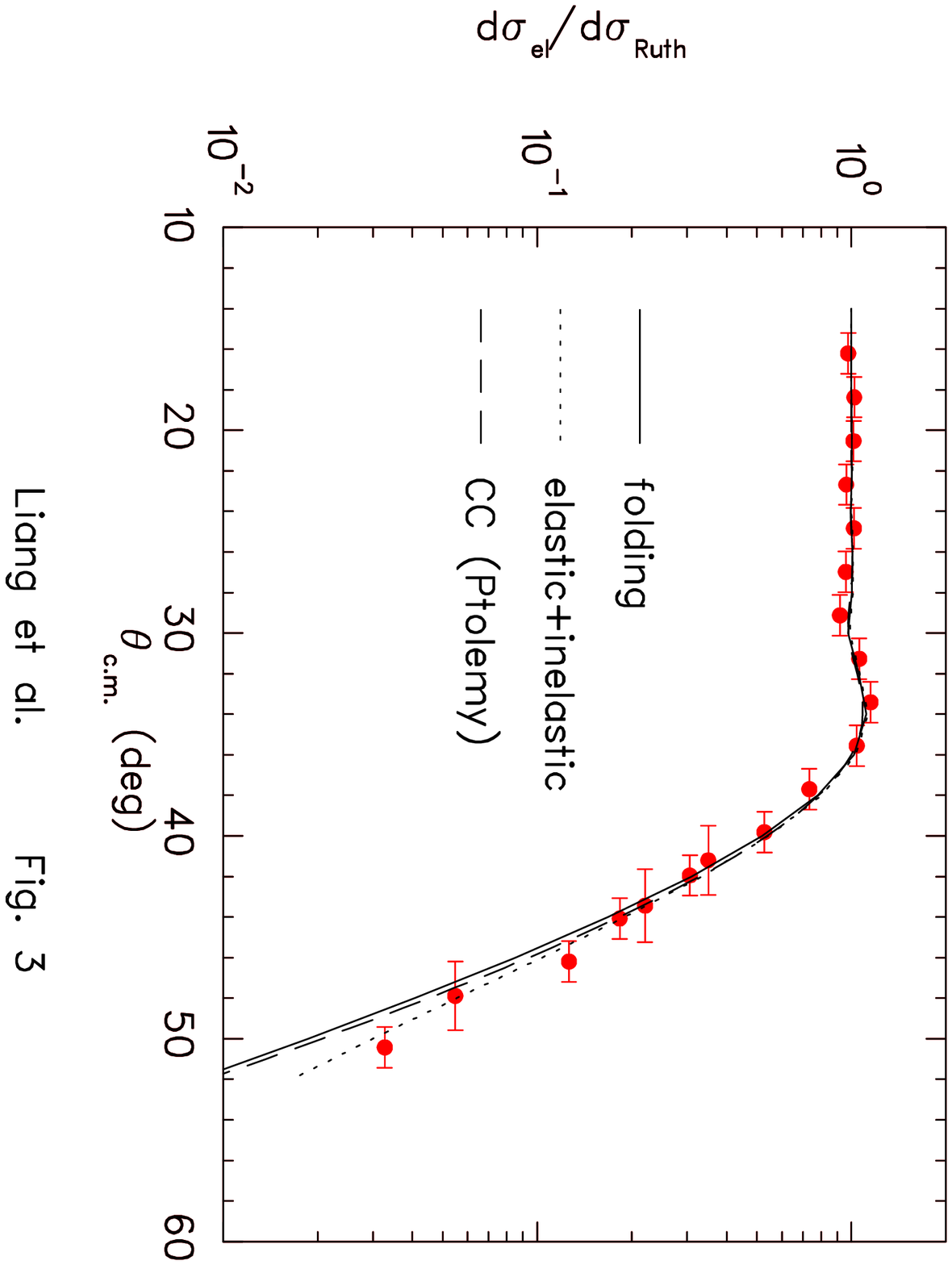}
\end{figure}
\begin{figure}
\epsfxsize=6.0in
\epsffile{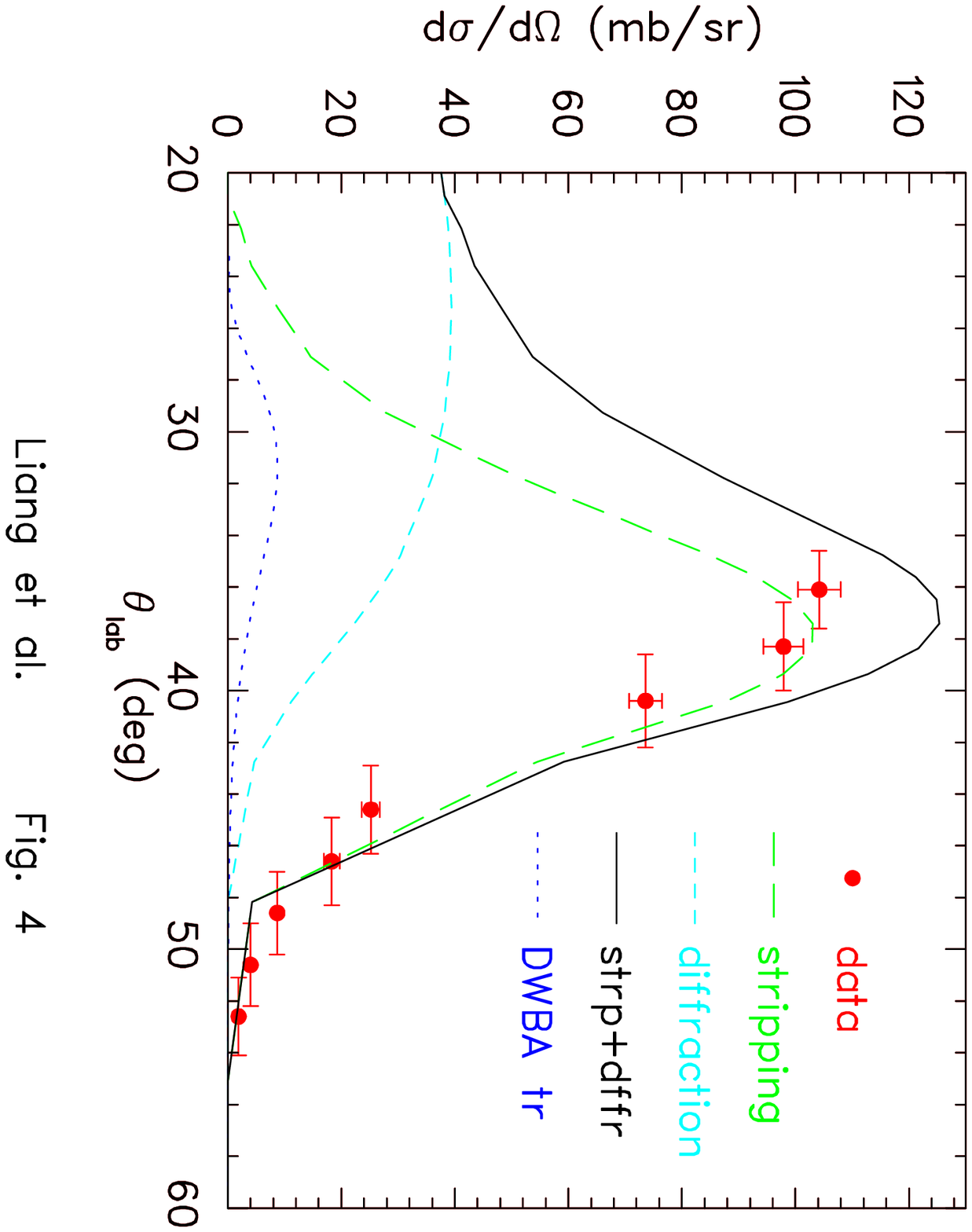}
\end{figure}
\begin{figure}
\epsfxsize=6.0in
\epsffile{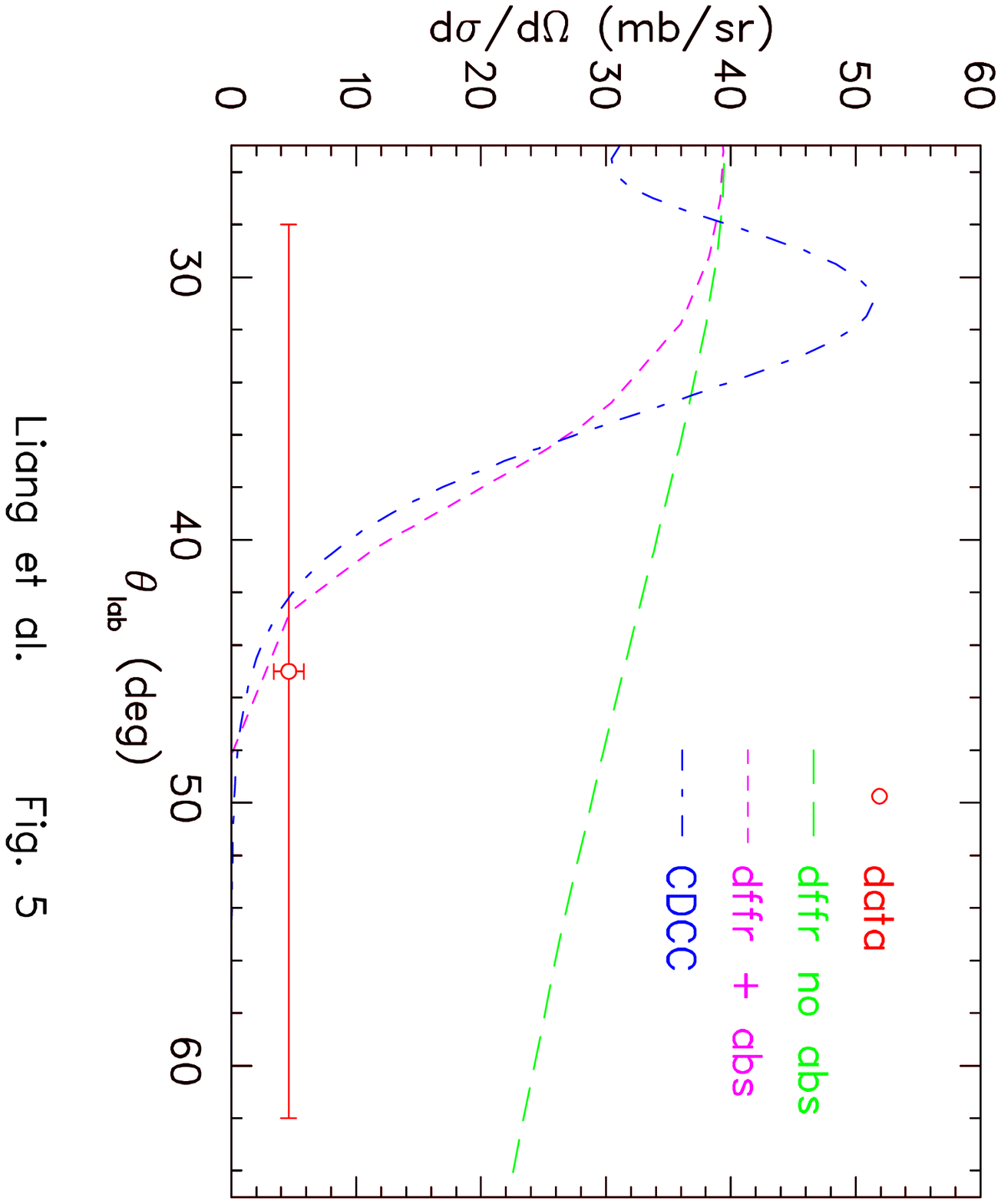}
\end{figure}

\end{document}